\documentstyle{aipproc2}

\def\ga{\gamma}

\def\si{\sigma}

\def\ps{\psi}
\def\om{\omega}

\def\De{\Delta}

\def\cE{{\cal E}}
\def\fr#1#2{{{#1} \over {#2}}}

\def\half{{\textstyle{1\over 2}}}
\def\frac#1#2{{\textstyle{{#1}\over {#2}}}}

\def\lsim{\mathrel{\rlap{\lower4pt\hbox{\hskip1pt$\sim$}}
    \raise1pt\hbox{$<$}}}
\def\gsim{\mathrel{\rlap{\lower4pt\hbox{\hskip1pt$\sim$}}
    \raise1pt\hbox{$>$}}}
\def\sqr#1#2{{\vcenter{\vbox{\hrule height.#2pt
         \hbox{\vrule width.#2pt height#1pt \kern#1pt
         \vrule width.#2pt}
         \hrule height.#2pt}}}}

\newcommand{\beq}{\begin{equation}}
\newcommand{\eeq}{\end{equation}}
\newcommand{\bea}{\begin{eqnarray}}
\newcommand{\eea}{\end{eqnarray}}
\newcommand{\rf}[1]{(\ref{#1})}

\begin{document}

\begin{flushright}
{COLBY 98-04\\}
{IUHET 393\\}
{August 1998\\}
\end{flushright}

\title{Testing CPT and Lorentz Symmetry with \\
Electrons and Positrons in Penning Traps\footnote
{\footnotesize Presented by R.B.\ 
at the 1998 Conference on Trapped Charged Particles
and Fundamental Physics,
Pacific Grove, California, August 1998}
}

\author{Robert Bluhm$^a$, V. Alan Kosteleck\'y$^b$, 
and Neil Russell$^b$}
\address{$^a$Physics Department, Colby College, Waterville, ME, 04901 U.S.A.\\
\smallskip
$^b$Physics Department, Indiana University,
Bloomington, IN, 47405 U.S.A.}

\maketitle

\begin{abstract}
We present a theoretical analysis of signals for CPT
and Lorentz violation in $g-2$ and charge-to-mass-ratio
experiments on electrons and positrons in Penning traps.
Experiments measuring anomaly frequencies are found to
be the most sensitive to CPT violation.
We find that the conventional figure of merit
for CPT breaking,
involving the difference of the electron and positron
$g$ factors,
is inappropriate in this context,
and an alternative is introduced.
Bounds of approximately $10^{-20}$ are attainable.
\end{abstract}

\section*{INTRODUCTION}

The CPT theorem 
\cite{cpt} 
is a general and powerful result
that holds for local relativistic quantum 
field theories of point particles 
in flat spacetime.
Any field theory of this kind must be invariant under 
the combined operations
of charge conjugation C, parity reversal P, 
and time reversal T.
As a consequence of this invariance,
particles and antiparticles have equal masses, lifetimes,
charge-to-mass ratios, and gyromagnetic ratios.  
The CPT theorem has been tested to great accuracy in
a variety of experiments
\cite{pdg}.
The sharpest bound is obtained in experiments with neutral kaons,
where the CPT figure of merit is
\beq
r_K \equiv \fr {|m_K - m_{\overline{K}}|} {m_K}
\lsim 2 \times 10^{-18}
\quad .
\label{rK}
\eeq

Experiments on electrons and positrons
confined in Penning traps also yield
sharp bounds on CPT violation.
Indeed,
these experiments provide the tightest bounds on CPT
in the lepton system.
Two types of experimental comparisons 
of electrons and positrons are possible in Penning traps.
They involve making accurate measurements of cyclotron
frequencies $\om_c$ and anomaly frequencies $\om_a$ of
single isolated particles confined in the trap.
The first compares the ratio $2 \om_a / \om_c$ for particles
and antiparticles.
In the context of conventional 
quantum electrodynamics,
this ratio equals $g-2$ for the particle or antiparticle.
A second experiment compares values of $\om_c \sim q/m$,
where $q>0$ is the magnitude of the charge and $m$ is the mass, 
and is therefore a comparison of charge-to-mass ratios.

The conventional figure of merit adopted 
in $g-2$ experiments on electrons and positrons
is given as the relative difference in their $g$ factors
\cite{vd87,gb86},
\beq
r_g^e \equiv 
\fr {|g_{e^-} - g_{e^+}|} {g_{\rm avg}}
\quad ,
\label{rg}
\eeq
which is known to be less than $2 \times 10^{-12}$.
The bound obtained in charge-to-mass-ratio experiments
\cite{svdd81}
is expressed as the ratio
\beq
r_{q/m}^e \equiv 
\fr {\left|({q_{e^-}}/{m_{e^-}})
- ({q_{e^+}}/{m_{e^+}})\right|} {(q/m)_{\rm avg}}
\quad ,
\label{rqme}
\eeq
which is less than or equal to $1.3 \times 10^{-7}$.

Measurements of frequencies in atomic systems typically have
experimental uncertainties four or five orders of magnitude
better than the measurements made in kaon experiments.
However,
the figure of merit $r_g^e$ is poorer than $r_K$ by
about six orders of magnitude.
This raises some interesting questions about the Penning-trap
experiments as to why they do not provide better tests
of CPT despite having better experimental precision.
However,
it is impossible to pursue these types of questions
in the context of conventional quantum electrodynamics,
since CPT breaking is strictly forbidden.
Instead,
one would need to work in the context of a theoretical
framework that incorporates CPT-violating interactions,
making possible an investigation of possible experimental
signatures.
Only recently has such a theoretical framework 
in the context of the standard model been developed
\cite{ck}.

In this paper,
we summarize the results of our analysis on CPT and Lorentz
tests performed with electrons and 
positrons in Penning traps.
A more complete description of this analysis
can be found in
Refs.\ \cite{bkr1,bkr2}.

\section*{THEORETICAL FRAMEWORK}

The theoretical framework we use 
\cite{ck}
is based on a general extension of the
SU(3) $\times$ SU(2) $\times$ U(1) standard model
in particle physics.
It includes all possible leading-order
CPT- and Lorentz-violating
interactions that could arise
from spontaneous symmetry breaking at a 
more fundamental level, 
such as in string theory.
This type of CPT violation is a possibility in string
theory because the usual axioms of the CPT theorem do
not apply to extended objects like strings.
In spontaneous symmetry breaking,
the dynamics of the action remains CPT invariant,
which means the framework can preserve desirable features
of quantum field theory
such as gauge invariance, power-counting renormalizability,
and microcausality.
The CPT and Lorentz violation occurs only in the
solutions of the equations of motion and is
similar to the spontaneous breaking of the electroweak theory in 
the standard model.

To analyze interactions involving electrons and positrons
in a Penning trap,
we use a restriction of the full
particle-physics framework to quantum
electrodynamics.
The resulting model divides into two sectors,
one that breaks CPT and one that preserves CPT,
while both break Lorentz symmetry.
Possible violations of CPT and Lorentz symmetry
are parametrized by quantities that can be bounded by experiments. 
Within this framework,
the modified Dirac equation describing a fermion
with charge $q$ and mass $m$ 
in an electromagnetic field is given by
\beq
\left( i \ga^\mu D_\mu - m - a_\mu \ga^\mu
- b_\mu \ga_5 \ga^\mu - \half H_{\mu \nu} \si^{\mu \nu} 
+ i c_{\mu \nu} \ga^\mu D^\nu 
+ i d_{\mu \nu} \ga_5 \ga^\mu D^\nu \right) \ps = 0
\quad .
\label{dirac}
\eeq
Here, 
$\ps$ is a four-component spinor,
$i D_\mu \equiv i \partial_\mu - q A_\mu$
is the covariant derivative,
$A^\mu$ is the electromagnetic potential in the trap,
and $a_\mu$, $b_\mu$, 
$H_{\mu \nu}$, $c_{\mu \nu}$, $d_{\mu \nu}$
are the parameters describing possible violations
of CPT and Lorentz symmetry.
The properties of $\ps$ under transformations
imply that the terms involving
$a_\mu$, $b_\mu$ 
break CPT
while those involving 
$H_{\mu \nu}$, $c_{\mu \nu}$, $d_{\mu \nu}$ 
preserve it,
and that Lorentz symmetry is broken by all five terms.

Since there have been no experimental observations to date 
of CPT or Lorentz breaking,
the quantities $a_\mu$, $b_\mu$, 
$H_{\mu \nu}$, $c_{\mu \nu}$, $d_{\mu \nu}$ 
must all be small.
We can estimate the suppression scale for these
parameters by taking the scale governing the
fundamental theory as the Planck mass $m_{\rm Pl}$
and the low-energy scale as the electroweak 
mass scale $m_{\rm ew}$.
The natural suppression scale for Planck-scale effects
in the standard model would then be of order
$m_{\rm ew}/m_{\rm Pl} \simeq 10^{-17}$.

\section*{EXPERIMENTS IN PENNING TRAPS}

We use this theoretical framework to analyze comparative tests of
CPT and Lorentz symmetry on electrons and positrons
in Penning traps.
First,
we note that the time-derivative couplings in
\rf{dirac} alter the standard procedure for obtaining
a hermitian quantum-mechanical hamiltonian operator.
To overcome this,
we first perform a field redefinition at the lagrangian
level that eliminates the additional time derivatives.
We then use charge conjugation to obtain a Dirac equation
and hamiltonian for the antiparticle.

In tests of CPT,
experiments compare the cyclotron and anomaly
frequencies of particles and antiparticles.
According to the CPT theorem,
electrons and positrons of opposite spin in a
Penning trap with the
same magnetic fields but opposite electric fields should
have equal energies.
The experimental relations $g-2 = 2 \om_a / \om_c$ and 
$\om_c = qB/m$ provide connections to the quantities
$g$ and $q/m$ used in defining the figures of merit
$r_g^e$ and $r_{q/m}^e$.
We perform calculations using 
Eq.\ \rf{dirac} to obtain possible shifts in the
energy levels due to either CPT-breaking or CPT-preserving
Lorentz violation.
In this way,
the effectiveness of Penning-trap
experiments on electrons and positrons
as tests of both CPT-breaking and
CPT-preserving Lorentz violation
can be analyzed.
From the computed energy shifts we determine how the
frequencies $\om_c$ and $\om_a$ are affected and whether
the conventional figures of merit are appropriate.

In experiments performed in Penning traps,
the dominant contributions to the energy come from
interactions of the electron or positron with the
constant magnetic field of the trap,
while the quadrupole electric fields generate smaller effects.
In a perturbative calculation,
the dominant CPT- and Lorentz-breaking effects can therefore
be obtained by working with relativistic Landau levels
as unperturbed states.
Conventional perturbations,
such as the usual corrections to the anomalous magnetic moment,
are the same for electrons and positrons.
Violations of CPT and Lorentz symmetry result in
either differences between electrons and positrons
or in unconventional effects
such as diurnal variations in the measured frequencies.

\section*{RESULTS}
 
The results of our calculations for electrons and positrons in
Penning traps 
\cite{bkr1,bkr2}
show that the leading-order effects
due to CPT and Lorentz breaking cause corrections to the
cyclotron and anomaly frequencies:
\beq
\om_c^{e^-} \approx \om_c^{e^+} \approx
(1 - c_{00}^e - c_{11}^e - c_{22}^e) \om_c
\quad ,
\label{wcelec}
\eeq
\beq
\om_a^{e^\mp} \approx \om_a
\mp 2 b_3^e + 2 d_{30}^e m_e + 2 H_{12}^e
\quad .
\label{waelec}
\eeq
In our notation,
$\om_c$ and $\om_a$ represent the
unperturbed frequencies for the
electron ($e^-$) and the positron ($e^+$),
while $\om_c^{e^\mp}$ and $\om_a^{e^\mp}$ denote
the frequencies including the corrections.
Superscripts have also been added on the coefficients
$b_\mu$, etc., to denote that these parameters
describe the electron-positron system.
From these relations we find the differences in the
electron and positron cyclotron and anomaly frequencies to be
\beq
\De \om_c^e \equiv \om_c^{e^-} - \om_c^{e^+} \approx 0
\quad , 
\label{delwc}
\eeq
\beq 
\De \om_a^e  \equiv \om_a^{e^-} - \om_a^{e^+} \approx - 4 b_3^e
\quad .
\label{delwa}
\eeq
We find that
in the context of this framework,
comparisons of cyclotron frequencies to leading order
do not provide a signal for CPT or Lorentz breaking,
since the corrections to $\om_c$ for electrons and
positrons are equal.
However,
comparisons of anomaly frequencies provide unambiguous tests of CPT
since the CPT-violating term with $b_3$ results in
a nonzero value for the difference $\De \om_a^e$,
while the CPT-preserving coefficients do not appear.

We also find that to leading order there are no
corrections due to CPT or Lorentz violation
to the $g$ factors for either
electrons or positrons.
This leads to some interesting and unexpected results
concerning the figure of merit $r_g$ in
Eq.\ \rf{rg}.
With $g_{e^-} \approx  g_{e^+}$ to leading order,
we find that $r_g$ vanishes,
which would seem to indicate the absence of CPT breaking.
However,
this conclusion would be incorrect
because the model contains explicit CPT violation.
In addition,
our calculations show that with $\vec b \ne 0$ the
experimental ratio $2 \om_a / \om_c$ depends on
the magnetic field and 
is undefined in the limit of a vanishing $B$ field.
Therefore,
the usual relation $g-2 = 2 \om_a / \om_c$ does not hold
in the presence of CPT violation.
For these reasons,
we conclude that
the figure of merit $r_g$ in
Eq.\ \rf{rg} is inappropriate in the context of our framework.
An alternative is suggested next.

Since a prediction of the CPT theorem is that 
electron and positron states of opposite spin 
in the same magnetic field have equal energies,
we propose as a model-independent figure of merit
\beq
r^e_{\om_a}
\equiv \fr{|{E}_{n,s}^{e^-} - {E}_{n,-s}^{e^+}|}
{{E}_{n,s}^{e^-} }
\quad ,
\label{re}
\eeq
where ${E}_{n,s}^{e^\mp}$ are the energies of the
relativistic states labeled by their Landau-level 
numbers $n$ and spin $s$.
Our calculations show $r^e_{\om_a} \approx 
| \De \om_a^e | / 2 m_e  \approx |2 b_3^e | / m_e$.
Assuming frequency resolutions on the order of 1 Hz,
we estimate as a bound on this figure of merit,
\beq
r^e_{\om_a} \lsim 10^{-20}
\quad .
\label{relim}
\eeq

This definition of the figure of merit $r^e_{\om_a}$
is compatible with the corresponding
figure of merit $r_K$ 
arising from experiments with the neutral-kaon system.
This is because both figures of merit involve ratios
of energy scales,
and therefore comparisons across experiments are
more meaningful.
This is not the case for the figures of merit
$r_g^e$ and $r_K$,
since each involves different physical quantities.
Our estimate suggests that a somewhat tighter bound 
for $r^e_{\om_a}$ is attainable in Penning-trap experiments
than that for the corresponding figure of merit $r_K$ 
arising from experiments with the neutral-kaon system.
This result is more in line with the greater precision 
that is experimentally accessible in frequency
measurements in a Penning trap.
However, performing the CPT tests in the 
kaon system remains essential
because neutral-meson CPT violation is controlled 
by distinct CPT-violating parameters that appear 
only in the quark sector
\cite{k98}.

In Ref.\ \cite{bkr2},
we describe additional possible signatures of CPT and Lorentz 
violation.
These include possible diurnal variations in the anomaly
and cyclotron frequencies.
Tests for these effects would provide 
bounds on some of the components of the
parameters $c_{\mu \nu}^e$, $d_{\mu \nu}^e$, and $H_{\mu \nu}^e$.

One type of experiment looking for diurnal variations involves the
electron alone or the positron alone.
In the standard-model extension,
these variations would occur because
the components of the couplings in 
Eq.\rf{waelec}
would change as the Earth rotates.
Consider the following quantities for the
electron and positron:
\beq
\De^{e}_{\om_a^{e^-}} 
\equiv 
\fr {|{\cE}_{0,+1}^{e^-} - {\cE}_{1,- 1}^{e^-}|}
{{\cE}_{0,-1}^{e^-}} 
\quad ,  \quad \quad 
\De^{e}_{\om_a^{e^+}} 
\equiv 
\fr {|{\cE}_{0,-1}^{e^+} - {\cE}_{1,+1}^{e^+}|}
{{\cE}_{0,+1}^{e^+}} 
\quad .
\label{Deleorpdnlom}
\eeq
Suitable figures of merit $r^e_{\om_{a}^-,\rm diurnal}$  
and $r^e_{\om_{a}^+,\rm diurnal}$  
can be defined as 
the amplitude of the diurnal variations 
in $\De^{e}_{\om_a^{e^-}}$
and $\De^{e}_{\om_a^{e^+}}$, 
respectively.
In the context of our framework,
we find
\beq
r^e_{\om_{a}^{\mp},\rm diurnal} 
\approx
\fr {2|\mp b_3^e + d_{30}^e m_e + H_{12}^e|}{m_e}
\quad .
\eeq
The experimental issues involved in obtaining a
bound on $r^e_{\om_{a}^{\mp},\rm diurnal}$ include 
maintaining stability in the magnetic field.
For example,
limiting variations in the magnetic field to a 
level of about 5 parts
in $10^9$ over the duration of the experiment
would keep any drift in the 
200 MHz anomaly frequency within a 1 Hz margin.
The data would also need to be suitably binned 
according to the orientation of the magnetic field 
as a function of star time.
A more elaborate approach to such diurnal experiments
would be to mount the apparatus on a suitable rotating platform
and thereby to investigate any geometrical dependence 
more directly.

An experiment of this nature on electrons alone
or positrons alone would 
bound the combination 
$\mp b_3^e + d_{30}^e m_e + H_{12}^e$
of couplings in the standard-model extension.
It would involve searching for leading-order corrections to the
anomaly and cyclotron frequencies 
which exhibit periodicities of approximately 24 hours.
Subleading order corrections involving tensor couplings
might exhibit 12-hour periodicities.
However, these effects would be suppressed relative
to the leading-order effects in Eq.\ \rf{Deleorpdnlom}.
All three of these quantities in Eq.\ \rf{Deleorpdnlom}
break Lorentz symmetry,
but only the coupling $b_3^e$ breaks CPT.
If a signal were detected,
it would indicate Lorentz violation
but not necessarily CPT violation.
It would provide strong motivation
for a subsequent experiment 
comparing anomaly frequencies of electrons and positrons,
which would bound the CPT-breaking parameter $b_3^e$
in isolation.

Data for this type of experiment on electrons alone already exist,
and a preliminary analysis has been performed
\cite{mit}.
Assuming a precision of approximately 1 Hz
in detecting diurnal variations, 
we estimate a bound on Lorentz breaking of
\beq
r^e_{\om_{a}^{\mp},\rm diurnal} \lsim 10^{-20}
\quad .
\eeq

\section*{CONCLUSIONS}

We find that the use of a general theoretical framework
incorporating CPT and Lorentz breaking allows a detailed
investigation of possible experimental signatures in 
Penning-trap experiments on electrons and positrons.
Our results indicate that the best tests of CPT symmetry
in Penning traps
emerge from comparisons of anomaly frequencies in $g-2$ experiments.
Our estimated bound on CPT  
from a variety of signals is approximately $10^{-20}$ 
in electron-positron experiments.
A table showing these estimated bounds is presented in 
Ref.\ \cite{bkr2}.
We also find that experiments searching for 
diurnal variations in electrons
alone can provide bounds on Lorentz breaking at a level
of approximately $10^{-20}$.

\section*{ACKNOWLEDGMENTS}
 
This work was supported in part
by the National Science Foundation 
under grant number PHY-9801869.


\begin{references}

\bibitem{cpt}
See, for example,
R.G.\ Sachs,
{\it The Physics of Time Reversal}
(University of Chicago Press, Chicago, 1987).

\bibitem{pdg}
See, for example,
R.M.\ Barnett et al.,
Review of Particle Properties,
Phys.\ Rev.\ D {\bf 54} (1996) 1.

\bibitem{vd87}
R.S.\ Van Dyck, Jr., P.B.\ Schwinberg, and H.G.\ Dehmelt, 
Phys.\ Rev.\ Lett.\ {\bf 59} (1987) 26;
Phys.\ Rev.\ D {\bf 34} (1986) 722.

\bibitem{gb86}
L.S.\ Brown and G.\ Gabrielse,
Rev.\ Mod.\ Phys.\ {\bf 58} (1986) 233.

\bibitem{svdd81}
P.B.\ Schwinberg, R.S.\ Van Dyck, Jr., and H.G.\ Dehmelt,
Phys.\ Lett.\ A {\bf 81} (1981) 119.

\bibitem{ck}
D. Colladay and V.A. Kosteleck\'y,
Phys.\ Rev.\ D {\bf 55} (1997) 6760;
Indiana University preprint IUHET 359,
Phys.\ Rev.\ D, in press (hep-ph/9809521). 

\bibitem{bkr1}
R. Bluhm, V.A. Kosteleck\'y and N. Russell,
Phys.\ Rev.\ Lett.\ {\bf 79} (1997) 1432.

\bibitem{bkr2}
R. Bluhm, V.A. Kosteleck\'y and N. Russell,
Phys.\ Rev.\ D {\bf 57} (1998) 3932.

\bibitem{kps}
V.A.\ Kosteleck\'y and S.\ Samuel,
Phys.\ Rev.\ Lett.\ {\bf 63} (1989) 224;
{\it ibid.},
{\bf 66} (1991) 1811;
Phys.\ Rev. D {\bf 39} (1989) 683;
{\it ibid.},
{\bf 40} (1989) 1886;
V.A.\ Kosteleck\'y and R.\ Potting,
Nucl.\ Phys.\ B {\bf 359} (1991) 545;
Phys.\ Lett.\ B {\bf 381} (1996) 89.

\bibitem{k98}
V.A.\ Kosteleck\'y,
Phys.\ Rev.\ Lett.\ {\bf 80} (1998) 1818.

\bibitem{mit}
R.\ Mittleman, private communication.

\end{references}
\end{document}